\documentclass[
aps,%
14pt,%
final,%
notitlepage,%
oneside,%
onecolumn,%
nobibnotes,%
nofootinbib,%
superscriptaddress,%
noshowpacs,%
centertags,
secnumarabic,]%
{extarticle}

\usepackage[headheight=17pt,headsep=7mm,left=18mm, right=18mm, top=18mm, bottom=18mm]{geometry}

\usepackage{amsfonts}
\usepackage{amssymb}
\usepackage{amsmath}
\usepackage{bm}
\usepackage{latexsym}

\newcommand{\beq}{\begin{eqnarray}}
\newcommand{\eeq}{\end{eqnarray}}
\newcommand{\be}{\begin{equation}}
\newcommand{\ee}{\end{equation}}
\def\la{\mathrel{\mathpalette\fun <}}

\def\fun#1#2{\lower3.6pt\vbox{\baselineskip0pt\lineskip.9pt
\ialign{$\mathsurround=0pt#1\hfil ##\hfil$\crcr#2\crcr\sim\crcr}}}

\newcommand{{\SD}}{\rm SD}

\newcommand{\ver}{\mbox{\boldmath${\rm r}$}}

\newcommand{\vep}{\bm p}
\newcommand{\veq}{\mbox{\boldmath${\rm q}$}}

\newcommand{\veS}{\mbox{\boldmath${\rm S}$}}
\newcommand{\veL}{\mbox{\boldmath${\rm L}$}}

\newcommand{\ves}{\mbox{\boldmath${\rm s}$}}

\newcommand{\ven}{\mbox{\boldmath${\rm n}$}}

\newcommand{\lan}{\langle}
\newcommand{\ran}{\rangle}

\title{The higher $\chi_{cJ}(nP)$, $h_c(nP)$ states and the role of the gluon-exchange potential}
\author{\textbf{A.~M.~Badalian}\\
NRC ``Kurchatov Institute" \\
Russia, Moscow}
\date{\today}

\begin{document}

\maketitle

\begin{abstract}
The masses, the fine-structure splitting, and two-photon decay widths of  the higher $nP$-charmonium states are calculated in the relativistic string model,
reduced to the spinless Salpeter equation, where the static potential has no fitting parameters and the $c-$quark mass has the physical value.
The resulting masses of $h_c(3P)$, $\chi_{c1}(3P)$, $\chi_{c0}(4P)$, $\chi_{c0}(5P)$, $\chi_{c1}(5P)$ are obtained in a good agreement with
the experimental masses of the LHCb resonances: $h_c(4300)$, $\chi_{c1}(4274)$, $X(4500)$, $X(4700)$, $X(4685)$. To test sensitivity of results to a chosen gluon-exchange
(GE) potential, three types of $V_{ge}(r)$ are considered. In first case the non-screened GE potential with large vector coupling at asymptotic, $\alpha_{\rm V}(\rm asym.)=0.635$,
and the $c-$quark mass $m_c=1.430$~GeV are taken; in second case a screened  $V_{ge}$ and $m_c=1.385$~GeV are investigated, and in third case  the GE potential is totally
suppressed, $V_{ge}=0$, and  $m_c=1.32$~GeV. The agreement with experiment is reached only if the same (universal) flattened confining potential, introduced
in the analysis of the radial Regge trajectories of light mesons, is used. The unobserved $6P,0^+$ resonance with the mass $\cong 4.826(13)$~GeV, near $J/\psi\phi(1680)$
threshold, is predicted. Our analysis shows that the screening of the GE potential is possible but weakly affects the physical results obtained.
The calculated two-photon decay widths weakly differ in the three cases but may become an important factor, which distinguishes $c\bar c$ and four-quark states.
\end{abstract}

\section{Introduction}

In 2016 the LHCb observed the $0^+$ resonances $X(4500),~ X(4700)$ in the $J/\psi \phi$ system \cite{1} and   recently $X(4500)$
was confirmed in new experiments \cite{2,3}, where also the resonance $X(4685)$ with $J^P = 1^+$ was seen. These resonances have called out a great interest
among theoreticians and were interpreted mostly as four-quark states  \cite{4}-\cite{8}. However, there rather small mass $M_4(1S)$ of the lowest tetraquark
state with the $J^{PC}=0^{++}$ was obtained, e.g. it is equal to 4356~MeV in \cite{4} or 4051 MeV in \cite{9}. On the contrary, far ago \cite{10}
in the diquark-antidiquark system ($cs\bar c\bar s$) the larger mass $M_4(1S)=4490$ MeV was predicted, which agrees with the mass of $X(4500)$.
Nevertheless, one cannot  exclude  that $X(4500), X(4700)$ may be conventional $\chi_{c0}(4P), \chi_{co}(5P)$ mesons, shifted down due to coupling to open channels.
Such shifts of the $nP$ excitations were already studied, using the so-called flattened confining potential (CP), introduced in the analysis
of the radial Regge trajectories of light mesons \cite{11}, and for  $\chi_{c0}(4P)$ and $\chi_{c0}(5P)$ a good agreement with the experimental masses of
$X(4500)$ and $X(4700)$ was found in \cite{12}; a good agreement with experiment was also obtained in the analysis of the radial Regge trajectories in \cite{13}.
Also the mass shifts are possible due to  admixture of the $c\bar c$ state and a meson-meson component \cite{14}  and in the coupled-channel models \cite{15,16}, while
in the quenched  approximation, if the linear CP is used, the masses of $\chi_{c0}(4P),~\chi_{c0}(5P)$  appear to be by $\sim (100-200)$~MeV larger than
those of the $X(4500), X(4700)$ resonances \cite{13,15}.

Recently new resonance $h_c(4300)$  was observed by the LHCb  \cite{17}) and  now five higher charmonium resonances are well established.  If they are considered
as the $nP$  states, then they have large sizes, $\sim (1.4-2.4)$~fm, and in this region the QCD dynamics is not fully understood up to now. Therefore below we will
investigate  conventional $nP~(n=3,4,5,6)$ mesons to compare their properties for non-screened and screened gluon-exchanged (GE) potentials.
At present the static potential $V_0(r)$ is well studied at the distances $r\la 0.50$ fm, where its  parameters  are known from the nonperturbative background
field theory \cite{18} and the lattice QCD \cite{19,20,21}. We remind here the main features of $V_0(r)$.

1. The QCD constant $\Lambda_{\overline{MS}}(n_f=3) = 320(20)$~MeV is defined  now with a good accuracy from the lattice studies \cite{19,20,21} and therefore
the vector QCD constant, $\Lambda_{\rm V}(n_f=3)=1.4753 \Lambda_{\overline{MS}}(n_f=3)= (472\pm 28)$~MeV is also known. Just the value $\Lambda_{\rm V}=500$~MeV will be
taken here. Also the exploited static potential $V_0(r)$  satisfies  the Sommer relation  for the force $F(r) = V_0'(r)$:   $r_0^2 F(r_0) = 1.65 $ at the point
$r_0 = 0.46$~fm \cite{22}.  To determine $\alpha_{\rm V}(r)$ at large distances we will use the infrared regulator $M_B = 1.15$~GeV, derived
in the background perturbation theory \cite{18,22}, and obtain  $\alpha_V(\rm asym.) =0.635$ at asymptotic,  which appears to be close to the phenomenological one-loop
coupling with small vector $\Lambda$, introduced far ago \cite{23}.

2. For  low charmonium states with their  r.m.s. $\la 1.0$~fm  we take the conventional value of the string tension, $\sigma_0 = 0.180(2)$~GeV$^2$,  known from
the analysis of the leading Regge trajectory in light mesons \cite{24}, while for the higher states of large sizes  we will take the  so-called flattened CP,
which imitates open channels.

Our calculations are performed below with the  relativistic string Hamiltonian (RSH) \cite{25,26}, where  there is no a fitting constant in the static potential,
often exploited in different potential models (PMs). This fact puts a restriction on the choice of the $c-$quark mass, which has to correspond to the chosen vector
coupling.  In particular, for strong vector coupling with $\alpha_{\rm V}(\rm asym.) = 0.60(4)$ the quark mass  $m_c$ has to be in narrow range $m_c=(1.43\pm 0.02)$~GeV.
Another advantage of the RSH is that it contains the physical quark masses: $m_q=0,~ m_s\sim 150$~MeV for a light and strange quarks, and the pole mass
$m_Q(\mu)$ for a heavy quark. This feature of the RSH is in contrast with other relativistic PMs, where large constituent quark masses, factually fitting parameters,
are used.

Long-distance behavior of a static potential is important to define the properties of the higher states, unfortunately, behavior of the  GE potential at large distances
is not studied yet on fundamental level. It is usually assumed that  $V_{ge}(r)$ is not screened at large $r$, although such screening of $V_{ge}$ is
possible, as it takes place with the confining potential (CP). The problem of screening of both GE and CP potentials was raised by P. Gonzalez in nonrelativistic screened model \cite{27},
where for $\chi_{c0}(4P)$ rather small mass, 4.325~GeV, was obtained. Possible screening of $V_{ge}(r)$ in four-quark systems was also discussed in \cite{28}.

On the contrary, the long-distance behavior  of the CP was studied in details, analyzing radial Regge trajectories of light and other  mesons
\cite{11,13,24}, where the CP  $V_f(r)= \sigma(r) r$ was shown to be flattening with decreasing string tension at large distances. In \cite{13}
the arguments were presented that such flattened CP can be considered as  universal and applied to all kinds of mesons, including charmonium. However,
this phenomenological  $\sigma(r)$ contains an additional parameter $R_0=(6.0\pm 1.0)$~GeV$^{-1} =1.2(2)$~fm, which demonstrates  that the coupled-channel effects
start  from the distances  $r\cong 1.2$~fm. Notice that  the CP $V_f(r)$  is very different from a screened CP $V_{scr}(r)$ \cite{29}, which is widely used  but with different
parameters (see \cite{15,29,30,31}  and references therein); this $V_{scr}$ is approaching to a constant at large $r$, thus making the quark and antiquark not confined.

In our paper we calculate the masses of the higher  charmonium $nP$ states and consider three variants to estimate the role of the GE potential:
(1) first, with non-screened $V_{ge}(r)$ and  $m_c = m_c(pole)=1.430$~GeV; (2) second, with screened  GE potential  and smaller
$m_c = 1.385$~GeV; (3) in third case   $V_{ge}=0$ and  $m_c=1.320$~GeV, close to that of the current mass,  $\bar m_c=1.275(25)$ ~GeV.  Also the  fine
structure splitting and two-photon decay widths of the higher $nP$ sates are calculated.

\maketitle

\section{The masses of  $\chi_{cJ}(nP)$ $(n=3,4,5,6)$ in a static potential with linear confining  term}

At first  we present the masses of the higher $nP$ states in a static potential $V_0(r)$  with linear CP plus the GE term,
\be
V_0(r)= \sigma_0 r - \frac{4\alpha_{\rm V}(r)}{3r},
\label{eq.01}
\ee
where the vector coupling $\alpha_{\rm V}(r)$ in the coordinate space is expressed via the coupling $\alpha_{\rm V}(q)$ in the momentum space,
\be
\alpha_{\rm V}(r) = \frac{2}{\pi} \int^\infty_0 {\rm d}q \, \alpha_{\rm V}(q) \, \frac{\sin(qr)}{q},
\label{eq.02}
\ee
The derivation of this coupling is given in Appendix A, using the Fourier transform of the potential $V_{ge}(r) = - C_{\rm F} \frac{\alpha_{\rm V}(r)}{r}$ in the coordinate
space through the perturbative potential $V_p(q) = - 4\pi C_{\rm F} \frac{\alpha_{\rm V}(q)}{q^2}$ in the momentum space \cite{32}, which is taken here in two-loop
approximation.

In calculations we use the following set of the parameters:
\begin{gather}
\sigma_0 = 0.180~{\rm  GeV}^2,~ \Lambda_{\rm V}(n_f=3) = 0.50~{\rm GeV},~ M_{\rm B} = 1.15~{\rm GeV}, \notag \\ \alpha_{\rm V}(\rm asym.)=0.635;~ m_c=1.430~ {\rm GeV}.
\label{eq.03}
\end{gather}
Here the conventional string tension  $\sigma_0 = 0.180$~GeV$^2$,  $\Lambda_V= 500$~MeV from the lattice data,  the infrared regulator $M_B=1.15$~GeV from the background
perturbation theory \cite{18} are taken. It gives the two-loop vector coupling  $\alpha_{\rm V}(\rm asym.)=0.635$ at asymptotics. The accepted $c-$quark mass,
$m_c(\rm one-loop)=1.430$~GeV, corresponds to the chosen coupling. Notice that in the RSH \cite{25,26} there is no a fitting constant in  $V_0(r)$ (\ref{eq.01}) and for that
reason the variation of  $m_c$ is very restricted.

With the parameters from (\ref{eq.03}) and linear CP the masses of low states, $J/\psi, \chi_{cJ}(1P)$, and $\psi(2S), \chi_{cJ}(2P)$), were  obtained
in a good agreement with experiment \cite{12,33}. However, the masses of the higher states $\chi_{c0}(4P),~\chi_{c0}(5P)$ turn out
to be by 150-200 MeV larger than those of the $X(4500),~X(4700)$ resonances; the same result is typical for other relativistic PM \cite{34}.

The  RSH $H(\rm str.)$, suggested in \cite{25,26} and developed later in \cite{24,33},
\be
H(\rm str.) = \omega(nL) + \frac{m_c^2}{\omega(nL)} + \frac{\vep^2}{\omega(nL)} + V(r),
\label{eq.04}
\ee
is expressed via the quark kinetic energy  $\omega(nL)$, which is determined  by the condition: $\frac{\partial H(\rm str.)}{\partial \omega} = 0$. It gives

\be
\omega(nL) = \sqrt{m_c^2 + \vep^2}.
\label{eq.05}
\ee
Then putting $\omega(nL)$ into (\ref{eq.04}), one arrives at the well-known spinless Salpeter equation (SSE),

\be
H(\rm SSE) = 2\sqrt{m_c^2 +\vep^2} + V(r),  ~~ (2\sqrt{m_c^2 + \vep^2} + V(r) )~\phi(r) = \tilde{M}_{cog} \phi(r),
\label{eq.06}
\ee
where the physical quark masses,  $m_q=0,~m_s\cong 150$~MeV (of light mesons), and $m_Q$, equal  to the pole quark mass, have to be taken.
This feature is in contrast with other relativistic PM, where large  constituent quark masses are used, being  actually
fitting parameters. In the SSE with linear + GE potential and the parameters (\ref{eq.03}) the kinetic energy $\omega(nP)$: 1.774, 1.831, 1.884, 1.935 (in GeV), found
for $3P,~4P,~5P,~6P$ states, is by $\sim (25-30)\%$ larger than $m_c$.

We pay attention at important property of the RSH in heavy quarkonia  - the eigenvalue of (\ref{eq.06}) is just equal
to the spin-averaged mass $\tilde{M}_{cog}(nP)$, which values together with the r.m.s. $r_n (nP) = \langle \sqrt{r^2}\rangle_{nP}$ are given in Table~\ref{tab.01}.

\begin{table} [h!]
\caption{The masses $\tilde{M}_{cog}(nP)$ (in MeV) and the r.m.s. $r_n(nP)$ (in fm) for the linear + GE potential $V_0(r)$}
\begin{center}
\label{tab.01}
\begin{tabular}{|c|c|c|}\hline
   State    & $\sqrt{<r^2>_{nP}}$ (in fm) &  $\tilde{M}_{cog}(nP)$  (in MeV)   \\ \hline

  $3P$           &  1.20                     &  4318  \\
  $4P$           &  1.43                     &  4642    \\
  $5P$           &  1.64                     &  4934   \\
  $6P$           &  1.84                     &  5201       \\ \hline
\end{tabular}
\end{center}
\end{table}
From the Table~\ref{tab.01} it is evident  that the spin-averaged masses $\tilde{M}_{cog}(4P),~ \tilde{M}_{cog}(5P)$ are by 150 MeV and  230~MeV larger than those
of the $0^+$ resonances $X(4500), X(4700)$. This result  is typical for different PMs,  if the linear $\sigma_0 r$ potential is used.
Moreover, $\tilde{M}(4P) = 4573$~ MeV,  $\tilde{M}(5P) = 4857$~MeV, remain large, even if GE potential $V_{ge}=0$ and small $m_c = 1.27$~GeV (equal to the current mass
$\bar m_c$) are taken. Thus one can conclude that in charmonium, as in light mesons, the higher states should be described either in many-channel
approaches, which are often  model-dependent, or taking a flattened universal CP from \cite{11,24}; we use this potential in next Sections.

\maketitle

\section{The higher charmonium states in static potential $V_{0f}(r)$ with a flattened CP and  non-screened GE potential}

Now in (\ref{eq.06}) we take the static potential $V_{0f}$,

\be
V_{0f}(r) = V_f(r) -\frac{4\alpha_{\rm V}(r)}{3 r},
\label{eq.07}
\ee
with the same non-screened  GE term with the parameters (\ref{eq.03}) and  the universal flattened CP $V_f$ from \cite{11,24},

\be
V_f(r) = \sigma(r) r, ~~\sigma(r) = \sigma_0 (1 - \gamma f(r)).
\label{eq.08}
\ee
Here $\sigma(r)$ is a decreasing function of $r$ and the parameter $\gamma = 0.40$  was defined with the 10\% accuracy \cite{11,24}.
The function $f(r)$,

\be
f(r) = \frac{\exp(r - R_0)}{ B_f + \exp(r - R_0)},
\label{eq.09}
\ee
includes two phenomenological parameters: $R_0 = 6.0$~GeV$^{-1}$ which shows that decreasing of $\sigma(r)$ starts at the distances $r > 1.0$ fm, and
the number $B_f = 15$, characterizing a scale of the string tension, e.g. for the $5P$ excitation  $\langle \sigma(r)\rangle_{5P} = 0.147$~GeV$^2$.
(Note we take  $\sigma_0=0.182$~GeV$^2$ to keep  $\langle \sigma(r)\rangle_{1P} = 0.180$~GeV$^2$ for the ground state).
Important feature of the SSE (\ref{eq.06}) is that in heavy quarkonia the eigenvalue is just equal to the spin-averaged mass $M_{cog}(nP)$,  which at the same time coincides
with the mass of $h_c(n\,^1P_1)$  with high accuracy. Here the spin-orbit and tensor interactions are considered perturbatively, in first approximation (see Appendix B).
The masses of $h_c(nP)$ and $\chi_{c0}(nP), \chi_{c1}, \chi_{c2}$ mesons are presented in Table~\ref{tab.02}, using the values of the spin-orbit and the tensor parameters,
given in Table~\ref{tab.07} in Appendix B. Notice that the values of the fine-structure (FS) splittings are rather small, $\sim (10-15)$~MeV.

\begin{table} [h!]
\small
\caption{The masses of the $nP~(n=3,4,5,6)$ charmonium states for the SSE with the static potential $V_{0f}(r)$ (\ref{eq.07}) with non-screened $V_{ge}$ and $m_c = 1.430$~GeV}
\label{tab.02}
\begin{center}
\begin{tabular}{|c|c|c|c|c|c|}\hline
state         & $M(h_c(nP))$          & $M(\chi_{c2}(nP))$ & $M(\chi_{c1}(nP))$ & $M(\chi_{c0}(nP))$ & exp. data \\ \hline

  $3P$           & 4284                   &  4299                 & 4275        & 4236             & $4294\pm 4^{+4}_{-3},~J^P=1^+$, \cite{2}\\
                 &                           &                &                 &                     &  $4307^{+6.4}_{-6.6}\phantom{`}^{+3.3}_{-4.1},~J^{PC}=1^{+-}$, \cite{17}\\
  $4P$          & 4550                    &   4564                 & 4542           &    4507         &  $4512.5^{+6.0}_{-6.2} \pm 3,~J^P=0^+$,   \cite{3} \\
                &                         &                        &                &                  & $4474\pm 3\pm 3$,    \cite{2}\\
  $5P$          & 4732                    & 4741             & 4726                 & 4701              &  $ 4704\pm 10^{+14}_{-24},~J^P=0^+$, \cite{1}  \\
                &                        &                   &                    &                  &  $4684\pm 7^{+13}_{16},~J^P = 1^+$, \cite{2}\\
  $6P$           &4876                    & 4886             &4870                   &    4844         &      \\\hline
\end{tabular}
\end{center}
\end{table}

From Table~\ref{tab.02} one can see that in the static potential with flattened CP plus non-screened GE potential ($\alpha_{\rm V}(\rm asym.)=0.635$
and $m_c=1.430$~GeV) the calculated  masses   $M(\chi_{c1}(3P) =4275$~MeV and  $M(h_c(3P)) = 4284$~MeV agree with experimental masses
of $\chi_{c1}(4274)$ \cite{2} and $h_c(4300)$ \cite{17}. Also  $M(\chi_{c0}(4P)) = 4507$~MeV and $M(\chi_{c0}(5P)) = 4701$~MeV  are obtained in precision agreement
with the LHCb data on the $X(4500), X(4700)$ \cite{1,2,3}. This result allows to conclude that these resonances can be  the $\chi_{c0}(4P),~ \chi_{c0}(5P)$ mesons, however,
one cannot exclude that their w.f.s have  not small admixture of a four-quark, or meson-meson components.

We pay also attention  that for this interaction  the order of levels  is standard:  $M(2^+) > M(1^+) > M(0^+)$ and  their sizes
 $r_n(4P)=1.70,~r_n(5P)=2.25,~ r_n(6P)=2.55$~(in fm) exceed  by (20--40)\% those in the static potential $V_0(r)$ with the linear CP.

\maketitle

\section{Possible screening of the GE potential}

In the last Section the masses of the higher $nP$ states were obtained in a good agreement with data, if the strong non-screened GE potential was taken.
Still it is of interest to look how the spectrum is changed for a screened GE potential. Such screening of the GE potential may be possible due to open channels, as it occurs with
the CP. The problem of screening of $V_{ge}$ was raised  in \cite{27}, but not solved yet. Here to clarify the situation we consider two variants:\\
Variant A, when $V_{ge}$  decreases as exponent at large $r > 1.0$~fm and  $m_c=1.385$~GeV;\\
Variant B, when $V_{ge}=0$ and  $m_c=1.320$~GeV, which value is close to the current mass $m_c=1.275(25)$~GeV \cite{20}.

In the case A
\be
V_{ge}(\rm scr.) = - \frac{4 \alpha_{\rm V}(r)}{3 r}~\exp(-\delta r),~~ \delta=0.20~{\rm GeV},
\label{eq.10}
\ee
decreases as the exponent with $\delta = 0.20$~GeV and at the point $R_0=1.20$~fm is by $\sim 70$~MeV smaller than without screening.
In the Table~\ref{tab.03} we give calculated masses of $h_c(nP)$ in both cases A and B, which in case B occur to be by $\sim (50-60)$~MeV smaller
for the $4P,5P$ states. For the $3P$ state $M(h_c(3P))$ is in agreement with the mass of new LHCb resonance $h_c(4300)$ \cite{17}.

\begin{table}
\caption{The masses of $h_c(nP)$ (in MeV) for the variants: A ($V_{ge}(\rm scr.)$ with $\delta = 0.20$~GeV, ~$m_c=1.385$~GeV) and
 B ($V_{0f} = V_f, ~V_{ge}=0, ~m_c=1.320$~GeV)}
\begin{center}
\label{tab.03}
\begin{tabular}{|c|c|c|c|}\hline

   mesone  & $M(h_c(nP))$, case A  & $M(h_c(nP))$, case B & exp. data \\ \hline

  $h_c(3P)$   &   4307      &  4289            & $ 4307^{+6.4}_{-6.6}`^{+3.3}_{-4.1}$,~\cite{17}         \\
  $h_c(4P)$    &  4559     &   4508      &       \\
  $h_c(5P)$    &  4726      &  4657    &       \\
  $h_c(6P)$   &  4868      &  4804      &      \\\hline

  \end{tabular}
  \end{center}
  \end{table}

The masses of the $0^+, 1^+, 2^+$ states are presented in Table~\ref{tab.04}, using the spin-orbit and tensor parameters, given in Appendix B. One can see
that in the case B the order of levels is changed, $M(0^+) > M(1^+) > M(2^+)$, because perturbative FS parameters $a_{so}^{(p)}(nP)=c_t(nP)=0$ for $V_{ge}=0$ and
$a_{np}(nP)$ is negative. For the screened $V_{ge}(\rm scr.)$ the order is kept standard and in both cases
$(M(\chi_{c0}(nP))~(n=4,5)$ agree with the masses of the $X(4500), X(4700)$ resonances \cite{1,2}. Also in the case B  $M(\chi_{c1}(5P))$ is in good agreement with
the mass of the $X(4685)$ resonance \cite{2}. We also predict the masses of the $6P$ states, which are not observed yet -- in the case B $M(\chi_{c0}(6P)) =4812$~MeV
is just near the $J/\psi\phi(1680)$ threshold, $M(\rm thres.)=4777$~MeV, while in case A it is by 27 MeV larger.

  \begin{table}
  \caption{The masses of the $\chi_{cJ}(nP)$ $(n=3,4,5,6)$ (in MeV) in the case A ($m_c=1.385$~GeV,~$\delta=0.20$~GeV) and  the case B ( $m_c=1.320$~GeV,~$V_{ge}=0$)}
  \begin{center}
  \label{tab.04}
  \begin{tabular}{|c|c|c|c|}\hline

  state             &   case A    &  case B  &     exp.data\\ \hline

  $\chi_{co}(3P)$ &    4258   &   4305        &             \\
  $\chi_{c0}(4P)$  & 4522      &  4520        &$ 4512^{+6.0}_{-6.2} \pm 30$ , \cite{3}\\
  $\chi_{c0}(5P)$  & 4700  &      4667            & $4704 \pm 10^{+14}_{-24},$  \cite{1} \\
  $\chi_{c0}(6P)$   & 4839    & 4812   &      \\

  $\chi_{c1}(3P)$   &4300      &  4297      & $4294 \pm 4^{+4}_{-3}$, \cite{2}     \\
  $\chi_{c1}(4P)$   & 4552      &   4514    &          \\
  $\chi_{c1}(5P)$   &4721      & 4662       &  $4684 \pm 7^{+13}_{-16}$, \cite{2} \\
  $\chi_{c1}(6P)$   &4863       & 4808      &     \\

  $\chi_{c2}(3P)$    &  4322     & 4282     &        \\
  $\chi_{c2}(4P)$     &  4570     & 4502    &   \\
  $\chi_{c2}(5P)$     &  4734     & 4652     &    \\
  $\chi_{c2}(6P)$      &  4877   & 4800   &     \\\hline

\end{tabular}
\end{center}
\end{table}

Thus we conclude - if  the same flattened CP is taken, then for $3P, 4P$ excitations in all three cases the  masses of the $0^+$, $1^+$, $2^+$, $1^{+-}$ states have rather close
values (difference $\la 30$ MeV), while for the $5P,~6P$ states the mass differences increase and can reach  $\sim 80$~MeV.
Note that for the higher $nP~(n\geq 4)$ excitations the relativistic kinematics becomes very important and in nonrelativistic PM, even if  a small string tension is
taken as in \cite{35}, calculated masses  would be by 100-150 MeV larger.

\maketitle

\section{Two-photon widths of $\chi_{c0}(nP)$, $\chi_{c2}(nP)$}

Here we calculate two-photon decay widths, which values can be important for several reasons, in particular, $\Gamma_{\gamma\gamma}(\chi_{c0}(nP))$
can be compared with two-photon decay widths of the diquark-antidiquark resonances, which were supposed to be rather small \cite{10}.
In \cite{10} for the scalar $\bar c\bar ccc$ and $\bar s\bar sss$ resonances the two-photon decay widths were estimated to be:
$\Gamma(\kappa_{4c}\rightarrow \gamma\gamma) \sim (0.1 - 1.0)$~eV and $\Gamma(\kappa_{4s}\rightarrow \gamma\gamma) \sim 30$~eV . From these two
values one can expect that the two-photon decay width of the $\bar c\bar scs$ scalar resonance has to be  between these two numbers:
$\Gamma(\kappa_{\bar c\bar scs}\rightarrow \gamma\gamma)\sim (10-15)$~eV.
This number is much smaller than the two-photon decay width of the $\chi_{c0}(4P)$ state (see below Table~\ref{tab.05}).

To define $\Gamma_{\gamma\gamma}(nP)$ one needs to know the derivatives of the wave functions (w.f.s)
at the origin $R'(0, nP)$, which, as well as the w.f.s $\phi(r)$ of the SSE (\ref{eq.06}), have irregular behavior near origin \cite{36} and have to be regularized.
Here for a regularization we use  the so-called einbein approximation (EA) to the RSH \cite{33,37}, which provides regular behavior of the w.f.s near origin and
as a whole the accuracy of the EA approximation was estimated to be around  5\%. Note that in the EA the kinetic
energy plays the role of effective quark mass and therefore the w.f. of the $nS$ state at $r=0$, as well as the derivative of the $nP$ w.f. at the origin,
are larger than those in non-relativistic case. For the  $nP$ state the ratio $\tilde{R}(r=0,nP) = \frac{R'(0,nP)}{\omega(nP)}$  plays the role of the w.f. at the origin.
Notice that in the flattened CP the  kinetic energies $\omega(nP)$ remain almost constant for the higher $nP$ states with $n=4,5,6$ and the derivatives $|R'(0,nP)|$
slowly decrease for the higher excitations, e.g. for non-screened $V_{ge}$ the ratios $\tilde{R}(0, nP)$ are following,

\begin{gather}
|\tilde{R}(0,3P)|=0.233~{\rm GeV}^{3/2},\,\, |\tilde{R}(0,4P)|=0.216~{\rm GeV}^{3/2}, \notag \\ |\tilde{R}(0,5P)|=0.206~{\rm GeV}^{3/2}.
\label{eq.11}
\end{gather}
These numbers $|\tilde{R}(0,nP)|$ occur to be only by $\sim 20\%$ larger  than corresponding
$|\tilde{R}(0,nP)| = 0.184, 0.178, 0.169$ (in GeV$^{3/2}$ in the case B with  $V_{ge}=0$.

Two-photon decay width of  a scalar $0^{++}$ meson \cite{33},
\be
\Gamma_{\gamma\gamma}(\chi_{c0}(nP)) = \frac{108~\alpha^2~e_c^4}{M_{\rm S}^2}~|\tilde{R}(0,nP)|^2 ~\eta_0,
\label{eq.12}
\ee
includes the perturbative correction $\eta_0 = 1 - \frac{28 - 3\pi^2}{9} \frac{\alpha_s}{\pi} =1.0111$  for  $\alpha_s = 0.195$ \cite{33}
and in the case B $\eta_0= 1.0$. This expression takes into account relativistic corrections through the w.f. $|\tilde{R}(0,nP)|$.

The two-photon decay width of a tensor meson \cite{33},

\be
\Gamma_{\gamma\gamma}(\chi_{c2}(nP))  = \frac{144~\alpha^2~ e_c^4}{5 ~M_{\rm T}^2} |\tilde{R}(0, nP)|^2 ~\eta_2, ~
\label{eq.13}
\ee
includes the perturbative factor  $\eta_2 = 1 - \frac{16}{3 \pi} \alpha_s = 0.669$  for  $\alpha_s=0.195$; in the case B $\eta_2=1.0$. Calculated
two-photon decay  widths are presented in  Table~\ref{tab.05}.

\begin{table}
\caption{The two-photon decay widths $\Gamma_{\gamma\gamma}(nP)$ of $\chi_{c0}(nP)$, $\chi_{c2}(nP)$ (in KeV); in the case B $\eta_0=\eta_2=1.0$}
\begin{center}
\label{tab.05}
\begin{tabular}{|c|c|c|c|} \hline

     Variants    & non-screened $V_{ge}$  &  Variant A   &  Variant B \\ \hline
    $m_c$ (in GeV)   & 1.430        &   1.385  & 1.320 \\

  $\chi_{c0}(3P)$   &2.82     &  2.63       & 2.08  \\
  $\chi_{c0}(4P)$   & 2.38     & 1.90        & 1.76  \\
  $\chi_{c0}(5P)$    &  1.72    & 1.44           &1.49  \\

  $\chi_{c2}(3P)$    &  0.48   &  0.45      & 0.56  \\
   $\chi_{c2}(4P)$   & 0.41    & 0.33      & 0.48  \\
   $\chi_{c2}(5P)$   &  0.30    &  0.25    &  0.39  \\\hline
\end{tabular}
\end{center}
\end{table}

As seen from Table~\ref{tab.05}, $\Gamma_{\gamma\gamma}(\chi_{c0})$, as well as $\Gamma_{\gamma\gamma}(\chi_{c2})$,  have close-by values for all three
variants, thus showing  that  the two-photon decay widths weakly depend  on the type of the GE interaction at large distances.
However,  an interesting result is obtained for the ratio  $R_{\gamma\gamma}(nP) = \frac{\Gamma_{\gamma\gamma}(\chi_{c0})}{\Gamma_{\gamma\gamma}(\chi_{c2})}$,
if one neglects a small difference between the masses of the scalar and the tensor mesons,  $M_{\rm S}(nP) \cong M_{\rm T}(nP)~(n=4,5)$. Then  the ratio,
\be
R_{\gamma\gamma} = \frac{\Gamma_{\gamma\gamma}(\chi_{c0}(nP))}{\Gamma_{\gamma\gamma}(\chi_{c2}(nP))} \cong  \frac{15}{4} =  3.75, ~
\label{eq.14}
\ee
if $\eta_0 =\eta_2=1.0$ (or $R_{\gamma\gamma}^{-1}=0.267$), when $V_{ge}=0$ or a contribution of the perturbative corrections is  neglected.

If the perturbative corrections  $\eta_0, \eta_2$ are important ($\eta_0=1.011,~\eta_2=0.669,~\alpha_s=0.195$), then  this ratio,
\be
 R_{\gamma\gamma} \cong \frac{15 \eta_0}{4 \eta_2} = 5.67,~~R_{\gamma\gamma}^{-1} = 0.176,
\label{eq.15}
\ee
becomes 1.5 times larger. This effect is not small and can be seen in  future experiments; it also allows
to estimate the role of perturbative effects in the higher charmonium states, which have large sizes, $\sim (1.80 - 2.5)$~fm.

\section{Conclusions}

In our paper we calculated the masses, the fine -structure splittings, and two-photon decay widths of the higher $nP$ charmonium states using
the RSH with three types of the GE potential and the same flattened CP. We have not found significant differences in the mass spectrum and this result means that
the dynamics of the higher excitations is mostly determined by the flattened CP. Comparison of predicted masses with the masses of five resonances, which are already
observed in experiments  --  $\chi_{c1}(4274)$, $h_c(4300)$, $X(4500)$, $X(4700)$, $X(4685)$ \cite{1,2,3,17}, demonstrates  good agreement for all five states, see Table~\ref{tab.06}.

\begin{table}
\caption{Comparison predicted masses (in MeV) of the $nP$ resonances with experimental data}
\begin{center}
\label{tab.06}
\begin{tabular}{|c|c|c|c|c|}\hline

Meson           & non-screened $V_{ge}$  &  Variant A  & Variant B  &experiment\\\hline
$m_c$~(in GeV)   &   1.430       & 1.385           &  1.320          &    \\

$\chi_{c1}(3P)$  &  4275         & 4301            & 4297           & $4294 \pm 4^{+4}_{-3}$, \cite{2}  \\

$h_c(3P)$        &  4284         & 4307           &  4289          & $4307^{+6.4}_{-6.6}`^{+3.3}_{-4.1}$, \cite{17}  \\

$\chi_{c0}(4P)$    & 4507        &4522             &  4520         & $4512^{+6.0}_{-6.2} \pm 30$, \cite{2} \\

$\chi_{c0}(5P)$   & 4701         & 4700          & 4667          &  $ 4704 \pm 10^{+14}_{-24}$, \cite{2}   \\

$\chi_{c1}(5P)$   &  4726        & 4718            & 4662         &  $4684 \pm 7^{+13}_{-16}$, \cite{2} \\\hline
\end{tabular}
\end{center}
\end{table}

From our point of view such coincidence cannot be occasional and  means that the $c\bar c$ component has to be an important part
of the w.f.s of these resonances, although this fact does not exclude a possible (not small)  admixture of a four-quark component.
We pay also attention that the sizes of $\chi_{cJ}(6P)$ states are not too large, $\sim 2.6(1)$~fm, -- they are just the same as the r.m.s.
of $\psi(4S)$, and therefore the resonances $\chi_{cJ}(6P)$ and $h_c(6P)$ can  exist. For $h_c(6P)$ we predict the mass $\cong 4.840(36)$~GeV,
close to the $J/\psi\phi(1680)$ threshold.

For three variants considered the values of  $\Gamma_{\gamma\gamma}(\chi_{c0}(nP))$ ($\Gamma_{\gamma\gamma}(\chi_{c2}(nP))$) differ only by (20-30)\%
for a given $nP$ state. However, their ratio  $R_{\gamma\gamma}$ (\ref{eq.15}) depends on
the perturbative corrections and changes from the number  $R_{\gamma\gamma}\cong 3.75~(n=4,5)$, if they are absent, up to the value $R_{\gamma\gamma}\cong 5.7$, if they
are taken into account. This difference can be observed  in future experiments.

For the $\chi_{c0}(4P)$ meson, considered as the pure $\bar cc$ state with the mass $M(\chi_{c0}(4P)=4515(8)$ MeV, the calculated
$\Gamma_{\gamma\gamma}(\chi_{c0}(4P))\sim (1.8 - 2.4)$~ KeV is not small and can be compared with the two-photon decay width of the $(0^+, \bar c\bar scs)$
diquark-antidiquark resonance, predicted in \cite{10} with the mass $M_4(1S)=4490$~MeV. In \cite{10} the estimate gives much smaller
$\Gamma_{\gamma\gamma}(0^+,\bar c\bar scs)\sim (10-15)$~eV. It means that the two-photon decay widths can be used as a good criterium to distinguish
between the higher $c\bar c$ states and tetraquarks. However, if the $\bar cc$ state is mixed with
the $\bar c\bar scs$ component, then its two-photon decay width will be smaller and depend on the mixing angle, which can be extracted from
future experimental data on the two-photon decay width of the $X(4500)$ resonance.

I am very grateful to Prof. Yu.~A.~Simonov for useful discussions and remarks.
I express my gratitude to Prof. B.~L.~G.~Bakker who has performed the calculations for our common papers some years ago, which were used here.

\section{Appendix A. {The vector coupling in the coordinate space}}

\setcounter{equation}{0} \def\theequation{A.\arabic{equation}}

The relation between the vector couplingg $\alpha_V(r)$ and $\alpha_V(q)$ in the momentum space (\ref{eq.02}) was derived in \cite{32}, using the Fourier transform of
the GE potential $V_{ge}(r) = - C_F \frac{\alpha_{\rm V}(r)}{r}$, where the GE potential in the momentum space is
$V_p(q) = - 4\pi C_F \frac{\alpha_{\rm V}(q)}{q^2}$,

\be
V_{ge}(r) = \int  {\rm d}^3 {q} \, \frac{V_p(q)}{(2\pi)^3} \, \exp(i\veq\ver).~~
\label{eq.A.1}
\ee
Performing integration over the angular variables one obtains

\be
\alpha_{\rm V}(r) = \frac{2}{\pi} \int^\infty_0 {\rm d}q \, \alpha_{\rm V}(q) \, \frac{\sin(qr)}{q}.
\label{eq.A.2}
\ee
This representation of the vector coupling was later used in many theoretical studies. In our approach the coupling $\alpha_{\rm V}(q)$ depends on the variable
$t_{\rm B}=\ln (\frac{q^2 + M_{\rm B}^2}{\Lambda_{\rm V}^2})$, where the IR regulator $M_{\rm B}$ is taken from
the background perturbation theory \cite{38}. This vector strong coupling has following properties: \\

First, as seen from (\ref{eq.A.2}), the asymptotic values  of the vector couplings $\alpha_{\rm V}(r)$ and $\alpha_{\rm V}(q=0)$ in the momentum space coincide:
$\alpha_{\rm V}(\rm asym.) = \alpha_{\rm V}(r \rightarrow \infty) = \alpha_{\rm V}(q=0)$. \\

Second, if in (\ref{eq.A.2}) the 1--loop  $\alpha_{\rm V}(q)^{(1)} = \frac{4\pi}{\beta_0 \ln t_{\rm B}}$ is taken with well established
QCD constant $\Lambda_{\rm V}(n_f=3) = 480(20)$~MeV  and $M_{\rm B}\cong 1.05(10)$~GeV, then the frozen value $\alpha_{\rm V}^{(1)}(q=0) = 0.82(2)$ appears to be
too large and cannot provide a good description of the heavy quarkonia spectra. On the contrary, the two-loop
$\alpha_{\rm V}^{(2)}(q) = \alpha_{\rm V}^{(1)}(q) \left( 1 - \frac{\beta_1}{\beta_0^2} \frac{\ln t_{\rm B}}{t_{\rm B}}\right) \cong 0.60(3)$ gives good description of the
charmonium spectrum.  Here $\beta_0= 11 - \frac{2}{3} n_f, ~ \beta_1 = 102 - \frac{38}{3} n_f$. \\

In \cite{22} the analysis of $\alpha_{\rm V}(r)$ was done with two different $\alpha_{\rm V}(q)$. In first case in Eq.~(\ref{eq.A.1}) the number of the flavors
$n_f=3$ in  $\alpha_{\rm V}(q)$ was taken at all momenta $q^2$ and two values of $\Lambda_{\rm V}(n_f=3)$ were studied:
$\Lambda_{\rm V}(n_f=3)=497$~MeV~($M_{\rm B}=1.15$~GeV) and  $\Lambda_{\rm V}(n_f=3) = 465$~MeV~ ($M_{\rm B}=1.10$~GeV).

In second case so-called compound coupling $\alpha_{\rm V}^{c}(r)$ was calculated through a compound coupling $\alpha_{\rm V}^{c}(q)$ in the momentum space, which
is defined in three different $q^2$-regions with corresponding $n_f$ and $\Lambda_{\rm V}(n_f)$:
\begin{enumerate}
\item{ at large $q \geq m_b, ~n_f=5, ~\Lambda_{\rm V}(n_f=5)=298(16)$ ~MeV \cite{20};}
\item in the region $ m_b \geq q \geq m_c,~ n_f=4, ~ \Lambda_{\rm V}(n_f=4) =421(14)$~MeV;
\item in the region $m_c \geq q \geq 0,~ n_f=3, ~\Lambda_{\rm V}(n_f=3) = 480(20)$~MeV were used.\\\end{enumerate}
Notice that chosen vector QCD constants $\Lambda_{\rm V}(n_f)$  correspond to well-established $\Lambda_{\overline{MS}}(n_f)=213(8), 296(10), 325(15)$~(in MeV)
for $n_f=5,4,3$, respectively \cite{19,20,21}.

The analysis \cite{22} has shown that  for both choices of $\alpha_{\rm V}(q)$  the vector coupling $\alpha_{\rm V}(r)$
practically coincides for all distances with exception of very small region, $r < 0.10$~fm. Therefore  $\alpha_{\rm V}(r)$, defined via
$\alpha_{\rm V}(q)$  with $n_f=3$, can be used for all mesons, including bottomonium, although the bottomonium ground state, which has a small size,
needs a small correction. In our paper $\alpha_{\rm V}(r)$ is defined via the two-loop  $\alpha_{\rm V}(q, n_f=3)$ with $n_f=3$ and in the variable
$t_{\rm B} = \ln \frac{q^2 + M_{\rm B}^2}{\Lambda_{\rm V}^2}$   well established vector $\Lambda_{\rm V}(n_f=3) =480(20)$~MeV is used.

\be
\alpha_{\rm V}(q) = \frac{4\pi}{9 t_{\rm B}} \left(1  -  0.790123 \frac{\ln t_{\rm B}}{t_{\rm B}}\right).
\label{eq.A.3}
\ee
In \cite{22} it was shown that different $\Lambda_{\rm V}(n_f=3)=497$ ~MeV ($M_{\rm B}=1.15$~GeV, $\alpha(\rm asym.)=0.635$) and
$\Lambda_{\rm V}(n_f=3) = 465$ ~MeV ($M_{\rm B}=1.10$~GeV, $\alpha_{\rm V}(\rm asym.)=0.6086$) give similar results.

\section{Appendix B. {The spin-orbit and tensor parameters}}

\setcounter{equation}{0} \def\theequation{B.\arabic{equation}}

The parameters of the spin-orbit potential,
\be
\hat{V}_{so}(r) = V_{so}(r) ~\veL\veS,
\label{eq.B.1}
\ee
include perturbative and nonpeturbative (NP) terms, $V_{so}(r)= V_{so}^{p}(r) + V^{np}(r)$, and for arbitrary $J, L,S$
\be
\veS\veL = \frac{1}{2} (J(J+1) -L(L+1) - S(S+1)).
\label{eq.B.2}
\ee
In the tensor potential,
\be
\hat{T} = V_t(r) \hat{S}_{12}, ~~ \hat{S}_{12} = 3((\ves_1\ven)(\ves_2\ven) - \ves_1\ves_2), ~\ven = \frac{\ver}{r},
\label{eq.B.3}
\ee
$V_t(r)$ is defined via the GE potential (see below) and

\be
\hat{S}_{12} =  - \frac{6(\veL\veS)^2 + 3\veL\veS - 2S(S+1)L(L+1)}{2(2L-1)(2L+3)}.~~\cite{39}
\label{eq.B.4}
\ee

The important feature of the RSH (\ref{eq.04}) is that the FS matrix elements $a_{so}(nP),~c_t(nP)$ are defined trough the
kinetic energy $\omega(nP)$, not via the mass $m_c$:
\be
a_{so}(nP)= \frac{\langle V_{so}(r)\rangle_{nP}}{\omega^2(nP)}, ~~c_t(nP) = \frac{\langle V_t(r)\rangle_{nP}}{\omega(nP)^2},
\label{eq.B.5}
\ee
thus taking into account relativistic kinematics \cite{24}. In the FS parameters we take  perturbative contributions in first approximation, since for the higher $nP$ states
second order corrections are very small, while for the ground state they can be important \cite{40}.

The tensor potential is defined by the  GE potential \cite{41},
\be
V_t(r) = - \frac{1}{r} \frac{\partial V_{ge}}{\partial r} = \frac{4}{3} \frac{\alpha_{fs}(r)}{r^3}, ~
\label{eq.B.6}
\ee
where the effective FS coupling is introduced,
\be
\alpha_{fs}(r) = \alpha_{\rm V}(r) - r \alpha_{\rm V}'(r). ~~
\label{eq.B.7}
\ee

It includes a negative term with the derivative $\alpha_{\rm V}'(r)$, which decreases the value of effective coupling by (10--15)\% \cite{22,32}.
The  m.e. $\langle r^{-3}\rangle_{nP}$  is calculated here in einbine approximation \cite{33,37}, since the w.f. of SSE has irregular behavior near origin.
In \cite{42} such regularization was not done and too large FS splittings were obtained.

Thus the tensor splitting is
\be
c_t(nP) = \frac{4 \alpha_{fs} \langle r^{-3}\rangle_{nP}}{3 \omega^2(nP)},~~
\label{eq.B.8}
\ee
As seen from (\ref{eq.B.8}), the scale of $\alpha_{fs}$ is mostly defined by the m.e. $\langle r^{-3}\rangle_{nP}$ and one can introduce the
characteristic distance  $r_{fs}(nP) = (\frac{1}{\langle r^{-3}\rangle_{nP}})^{1/3}$; its values are given in Table~\ref{tab.07}; they belong
to the range (2.2-2.8)~GeV$^{-1}$, where the vector coupling in the coordinate space was studied in details \cite{32}. Here for the $3P$ state $r_{fs}(3P)=2.2$~GeV$^{-1}$
and we take $\alpha_{fs}\cong 0.38$, which is by 10\% smaller than $\alpha_{\rm V}(r_{fs})\cong 0.42$,
due to negative derivative in (\ref{eq.B.7}). For the $nP$ states with $n\geq 4$ the distances $r_{fs}$ have close values, $r_{fs}=2.62(16)$~GeV$^{-1}$ and
one can take the same $\alpha_{fs}(nP)=0.44,~(n=4,5,6)$.

In considered approximation the spin-orbit splitting is
\be
a_{so}(nP) = \frac{3}{2} c_t(nP) + a_{np}(nP), ~~a_{np} = - \frac{\sigma \lan r^{-1}\ran_{nP}}{2\omega^2}.~~
\label{eq.B.9}
\ee
For the potential $V_{0f}$ (\ref{eq.07}) with non-screened GE potential the parameters $a_{so}, ~a_{np},~c_t$
are given in Table~\ref{tab.07}, they define the masses $M(n\,^3P_J)$:

\begin{gather}
M(n\,^3P_2)=M_{cog} + a_{so} - 0.1 c_t,~~M(n\,^3P_1)=M_{cog} - a_{so} + 0.5 c_t, \notag \\ M(n\,^3P_0)= M_{cog} - 2 a_{so} - c_t, .
\label{eq.B.10}
\end{gather}
which are given in Table~\ref{tab.02}.
\begin{table}
\caption{The FS parameters $a_{so}(nP),~a_{np}(nP),~c_t(nP)$ (in MeV) and $r_{fs}(nP)$ (in GeV$^{-1}$) for the potential $V_{0f}$ (\ref{eq.07}) with non-screened $V_{ge}(r)$ (\ref{eq.03})}
\begin{center}
\label{tab.07}
\begin{tabular}{|c|c|c|c|c|}\hline

state & $r_{fs}$ & $c(nP)$ & $a_{np}(nP)$  & $a_{so}(nP)$ \\

$3P$&    2.21    & 15,3      &  - 6.8        & 16.2    \\

$4P$   & 2.46     & 13.3   &   -5.0         & 14.9    \\

$5P$   & 2.78     & 9.7    &   -3.9        &   10.6 \\

$6P$   & 2.73     & 9.9      &  -3.6  &    11.2   \\\hline
\end{tabular}
\end{center}
\end{table}

For the screened GE potential (\ref{eq.10}) (the case A) all FS parameters are defined with $\alpha_{fs}=0.380$  and the following their values
are obtained:  $c_t(nP) = 15.9, 11.9, 8.5, 9.9$ and $a_{so}(nP) = 16.8, 12.5, 9.0, 9.9 ~(n=3,4,5,6)$ (in MeV); we estimate the uncertainties in
their numbers equal to $\pm 2$~MeV.  One can see that for the $V_{0f}$ potential with the screened $V_{ge}$ term the FS splittings are close to those,
given in Table~\ref{tab.07} for the potential with non-screened $V_{ge}$.  In the case A the masses of the $h_c(nP)$ and of the $\chi_{cJ}(nP)$ mesons are
given in Tables~\ref{tab.03} and \ref{tab.04}, respectively.

In the case B, when $V_{ge}=0$, the perturbative parameters $c_t(nP), a_{so}^{p}$ are equal to zero and the FS splittings are defined by
the negative $a_{np}$ (\ref{eq.B.9}), they are equal to  -- 8 , -- 6 , -- 5 , -- 4 ~$(n=3,4,5,6)$ (in MeV); therefore in this case the order of levels is changed:
$M(0^+) >  M(1^+) > M(2^+)$. The masses of the $\chi_{cJ}(nP)$ states in the case B are given in Table~\ref{tab.04}.

\end{document}